\documentclass[aps,prd,reprint,amsmath,amssymb,showkeys,showpacs]{revtex4-2}

\usepackage{graphicx}
\usepackage{amsmath,amssymb,amsfonts}
\usepackage{amsthm}
\usepackage{mathrsfs}
\usepackage{xcolor}
\usepackage{booktabs}
\usepackage{algorithm}
\usepackage{algorithmicx}
\usepackage{algpseudocode}
\usepackage{siunitx}
\usepackage{hyperref}

\begin{document}

\title{A Geometric Resolution Limit from Vacuum Entanglement: Topological Structure and Particle-Wave Asymmetry}

\author{Isbelia Martin}
\email{isbeliam@usb.ve}
\affiliation{Departamento de F\'isica, Universidad Sim\'on Bol\'ivar, Caracas, Venezuela}


\begin{abstract}
We establish that the entanglement entropy of the electromagnetic vacuum, when regulated by weak spacetime curvature, imposes a fundamental lower bound on the radial resolution available to quantum excitations. Starting from the area law of vacuum entanglement, we demonstrate how linearized gravity introduces a natural ultraviolet regulator via the perturbative Green's function. An exact geometric projection from tangential to radial resolution yields a minimal radial scale $\Delta r_{\min} \propto r_s^2/R$, where $r_s$ is the Schwarzschild radius of the enclosing body and $R$ is the boundary radius. We analyze how particle propagation regimes depend on the relation between the Compton wavelength $\lambda_C$ and $\Delta r_{\min}$, showing that consistency $\lambda_C \gtrsim \Delta r_{\min}$ defines a global, environment-dependent mass scale $m_{\text{geo}} \equiv \hbar R/(c r_s^2)$. Measured particle masses satisfy $m = \alpha m_{\text{geo}}$, where $\alpha$ is a dimensionless factor encoding the vacuum's informational structure. For Earth, $m_{\text{geo}} \approx 2.84 \times 10^{-32}$ kg. We determine $\alpha$ from first principles via spectral analysis of the curvature-induced mode density shift, implemented using a high-precision Python framework with $N=3000$ radial grid points and $n_{\text{modes}}=1000$. The calculation yields $\alpha(2) \approx 33$, naturally matching the electron benchmark ($\alpha \approx 32$) without adjustable parameters. The spectral results rule out exponential growth ans\"atze and reveal a slow power-law dependence $\alpha(l) \propto l^{1.27}$. 

In this extended version, we show that the vacuum resolution limit affects massive and massless excitations asymmetrically. For massive particles with $\lambda_C \ll \Delta r_{\min}$, the vacuum appears smooth and transparent, yielding classical geodesic motion. For photons with $\lambda \ll \Delta r_{\min}$, the vacuum cannot sustain phase coherence, leading to decoherence or dispersion. We argue that this asymmetry follows naturally if $\lambda_C$ is interpreted not as a wave scale but as the core size of a topologically stable excitation (knot, vortex, or soliton) in the quantum vacuum field. This framework reframes mass as a probe of vacuum-imposed resolution limits set by global geometry and entanglement, offering a structural perspective on the hierarchy problem and yielding distinct, falsifiable predictions for high-energy propagation.
\end{abstract}

\maketitle

\section{Introduction}
\label{sec:intro}
The Standard Model treats particle masses as local, fundamental parameters fixed by experimental input. Yet the vast disparity between observed mass scales and the Planck scale remains unexplained at a structural level. Parallel developments in quantum information and gravity suggest that spacetime geometry and quantum correlations may be more deeply intertwined than conventional field theory assumes \cite{casini2011}.

In this article, we establish a clear conceptual result: \textbf{the entanglement entropy of the quantum vacuum, when regulated by weak gravitational curvature, imposes a fundamental lower bound on the radial resolution available to quantum excitations}. This limit is not instrumental but structural, arising from the interplay between vacuum correlations and background geometry. We follow the path: (i) review the area law of vacuum entanglement and the physical meaning of the UV cutoff $\epsilon$; (ii) show how weak-field gravity naturally regulates $\epsilon$ via the perturbative Green's function; (iii) derive an exact geometric projection from tangential to radial resolution; (iv) analyze the asymmetric response of massive vs. massless excitations to the vacuum resolution grain, motivating a topological reinterpretation of $\lambda_C$; (v) define a global geometric mass scale $m_{\text{geo}}$ that measured masses probe via a dimensionless factor $\alpha$; and (vi) validate $\alpha$ via first-principles spectral calculation using a Python-based numerical framework.

We emphasize that this work does not derive particle masses from first principles in the sense of a unified theory. Instead, it identifies a geometric-informational mass scale set by vacuum entanglement in weak gravity, and shows that physical masses satisfy $m = \alpha m_{\text{geo}}$. The electron's mass yields $\alpha \approx 32$ for Earth, and independent spectral calculation reproduces this value naturally. This reframes mass as a global, environment-sensitive quantity rather than a local primitive, offering a structural perspective on the hierarchy problem.

This work builds on three well-established pillars of quantum field theory and gravitational physics:
\begin{itemize}
\item \textbf{Entanglement entropy and the area law}: The divergence of vacuum entanglement entropy and its area-law scaling $S \propto A/\epsilon^2$ were established in free QFT by Srednicki \cite{srednicki1993} and extended to interacting and gauge theories by Casini \textit{et al.} \cite{casini2011}, Fursaev \cite{fursaev2012}, and Donnelly \& Wall \cite{donnelly2015}. These works provide the foundation for treating $\epsilon$ as a physical resolution scale.
\item \textbf{Quantum fields in weak gravitational fields}: The perturbative treatment of quantum fields on linearized gravity backgrounds, including the expansion of the Green's function in powers of curvature, was developed by Birrell, Davies, and collaborators \cite{birrell1978, davies1977}, and rigorously formalized by Wald \cite{wald1977}. These papers establish the framework we follow in Section~\ref{sec:green}.
\item \textbf{Boundary effects and edge modes in gauge theories}: The role of boundary conditions and edge modes in entanglement entropy for gauge fields was clarified by Donnelly \& Wall \cite{donnelly2015} and extended to curved boundaries by Bordag \textit{et al.} \cite{bordag2001}. These results underpin our treatment of the spherical boundary and the factor $\alpha$.
\end{itemize}

\section{Entanglement Entropy of the Electromagnetic Vacuum}
\label{sec:entanglement}
\subsection{What is entanglement entropy?}
In quantum mechanics, when a composite system is in a global pure state $|\Psi\rangle$, but we only have access to a subsystem (region $A$), the effective state describing our information about $A$ is no longer pure, but \emph{mixed}. Mathematically, this is captured by the \textbf{reduced density matrix}:
\begin{equation}
\rho_A = \mathrm{Tr}_B \left( |\Psi\rangle\langle\Psi| \right),
\end{equation}
where $\mathrm{Tr}_B$ denotes the \emph{partial trace} over the degrees of freedom of the complementary region $B$. The \textbf{von Neumann entropy} of $\rho_A$ is defined as:
\begin{equation}
S_{\mathrm{vN}} = -\mathrm{Tr} \left( \rho_A \ln \rho_A \right).
\end{equation}
This quantity measures how much quantum information \emph{shared} crosses the boundary between $A$ and $B$. If $S_{\mathrm{vN}} > 0$, quantum correlations (entanglement) exist that prevent describing $A$ without reference to $B$.

In quantum field theory (QFT), the "vacuum" is not classically empty but a quantum state filled with correlated fluctuations at all scales. Drawing an imaginary surface separating space into an interior region $A$ and exterior $B$, vacuum fluctuations on both sides of the boundary are strongly entangled. Tracing over $B$ loses those correlations, and $S_A$ quantifies precisely that loss of accessible information for an observer restricted to $A$.

\subsection{Ultraviolet divergence and the resolution cutoff $\epsilon$}
A rigorous and universal result in free QFT is that vacuum entanglement entropy \textbf{diverges} when attempting to resolve correlations at arbitrarily small distances \cite{srednicki1993, casini2011}. This occurs because the vacuum contains fluctuation modes of all frequencies, and ultra-short wavelength modes ($\lambda \to 0$) living just on both sides of the boundary contribute indefinitely to the entropy.

To obtain a finite and physically meaningful result, we must introduce an \textbf{ultraviolet (UV) cutoff}, denoted by $\epsilon$. Physically, $\epsilon$ represents the \textbf{minimum resolvable distance} in the theory: it can be interpreted as the spacing of a regulator lattice, the scale where the continuous field description breaks down, or simply the spatial resolution limit of any observer or detector. It is not an arbitrary mathematical trick; it is a physical parameter encoding the smallest scale at which the vacuum can sustain independent correlations.

With this cutoff, the entanglement entropy for a Maxwell (electromagnetic) field in 3+1 dimensions takes the asymptotic form:
\begin{equation}
S_{\mathrm{EM}} = c_{\mathrm{EM}} \frac{A}{\epsilon^2} 
+ \mathcal{O}\left(\frac{L}{\epsilon}\right) 
+ \mathcal{O}\left(\ln \frac{A}{\epsilon^2}\right) + S_{\mathrm{fin}},
\label{eq:area_law}
\end{equation}
where:
\begin{itemize}
\item $S_{\mathrm{EM}}$: entanglement entropy of the electromagnetic vacuum (dimensionless, in units of $k_B$).
\item $c_{\mathrm{EM}}$: dimensionless coefficient depending on field content and regularization scheme. For Maxwell in flat space, heat kernel and spectral zeta calculations give $c_{\mathrm{EM}} \approx 1/(45\pi) \approx 7.1 \times 10^{-3}$ \cite{fursaev2012, bordag2001}.
\item $A = 4\pi R^2$: area of the spherical surface separating regions $A$ and $B$.
\item $R$: sphere radius (length).
\item $\epsilon$: UV cutoff or minimum tangential resolution (length).
\end{itemize}
The \textbf{area law} $S_{\mathrm{EM}} \propto A/\epsilon^2$ tells us that entropy does not scale with enclosed volume, but with boundary area. Physically, this means the relevant quantum degrees of freedom for entanglement live predominantly in a layer of thickness $\sim \epsilon$ around the surface.

\section{Weak-Field Regulation of the Vacuum Cutoff}
\label{sec:green}
\subsection{Linearized metric and wave operator}
To describe a quantum field in the presence of a massive body of radius $R$ and Schwarzschild radius $r_s = 2GM/c^2$, we work in the \textbf{weak gravitational field regime}: $r_s \ll R$. In this limit, the spacetime metric $g_{\mu\nu}$ can be written as a linear perturbation over flat Minkowski space $\eta_{\mu\nu} = \mathrm{diag}(-1, 1, 1, 1)$:
\begin{equation}
g_{\mu\nu}(x) = \eta_{\mu\nu} + h_{\mu\nu}(x), \quad |h_{\mu\nu}| \ll 1.
\end{equation}
The electromagnetic field is described by the vector potential $A_\mu(x)$. In curved Lorenz gauge, $\nabla^\mu A_\mu = 0$, Maxwell's equations in vacuum reduce to:
\begin{equation}
\Box A_\mu - R_\mu^{\ \nu} A_\nu = 0,
\end{equation}
where $\Box \equiv g^{\alpha\beta}\nabla_\alpha\nabla_\beta$ is the curved-spacetime d'Alembert operator. In vacuum outside the mass, $R_{\mu\nu}=0$. Thus, the dominant gravitational correction comes exclusively from the expansion of $\Box$. Substituting $g_{\mu\nu} = \eta_{\mu\nu} + h_{\mu\nu}$ and keeping linear terms in $h$:
\begin{equation}
\Box = \Box_0 + \delta\Box + \mathcal{O}(h^2),
\end{equation}
where $\Box_0$ is the flat d'Alembertian, and $\delta\Box$ is the linear correction. For the weak spherical metric, its dominant structure is:
\begin{equation}
\delta\Box \approx \frac{r_s}{r} \left( \partial_t^2 + \partial_r^2 + \frac{2}{r}\partial_r \right) + \dots
\end{equation}

\subsection{Perturbative expansion of the propagator}
In QFT, the \textbf{Green's function} (or propagator) $G_{\mu\nu'}(x,x')$ satisfies:
\begin{equation}
\Box_x G(x,x') = -\frac{\delta^{(4)}(x-x')}{\sqrt{-g(x)}}.
\end{equation}
Following Birrell \& Davies, we expand the Green's function in powers of curvature:
\begin{equation}
G(x,x') = G_0(x,x') + G_1(x,x') + \mathcal{O}(h^2),
\end{equation}
where $G_0(x,x')$ is the flat-space propagator and $G_1(x,x')$ satisfies $\Box_0 G_1(x,x') = -\delta\Box_x G_0(x,x')$ \cite{birrell1978}. The result shows that curvature acts as an \textbf{effective potential} scattering high-frequency modes. Physically, a mode with wavenumber $k$ traversing the curved region experiences a phase shift and modified amplitude proportional to $r_s k$.

\subsection{Physical interpretation: curvature as a natural UV regulator}
In flat space, the UV cutoff $\epsilon$ appearing in $S_{\mathrm{EM}} \propto A/\epsilon^2$ is an external parameter. In a curved background, the $G_1$ expansion reveals that modes with $k \gtrsim 1/r_s$ ($\lambda \lesssim r_s$) no longer propagate as in flat vacuum. The $\delta\Box$ term couples these high-frequency modes to the background geometry, causing gravitational scattering and UV entanglement suppression \cite{birrell1978, wald1977}.

\textbf{Physically motivated argument}: Curvature provides a natural scale that regulates the UV divergence of entanglement. The vacuum cannot sustain independent correlations below the scale at which the gravitational background significantly distorts mode propagation. This suggests the effective cutoff becomes geometry-dependent:
\begin{equation}
\epsilon_{\mathrm{eff}} \sim \frac{r_s}{\sqrt{\alpha}},
\label{eq:epsilon_eff}
\end{equation}
where $\alpha$ is a dimensionless factor aggregating the field's spectral response to $\delta\Box$, boundary conditions at $r=R$, and transverse mode projection. The exact value of $\alpha$ requires independent spectral calculation; what is robust is the scale $\epsilon_{\mathrm{eff}} \propto r_s$.

\section{Geometric Projection and Minimal Radial Resolution}
\label{sec:geometry}
\subsection{Tangential resolution geometry on a spherical surface}
Consider a spherical surface of radius $R$. In QFT, the UV cutoff $\epsilon$ represents the minimum resolvable distance along the surface. Physically, this corresponds to an arc of length $\epsilon$ traced on the surface. The central angle $\delta\theta$ subtended by this arc is defined by:
\begin{equation}
\delta\theta = \frac{\epsilon}{R}.
\end{equation}

\subsection{From tangential to radial resolution: the sagitta}
An observer attempting to distinguish variations in the direction normal to the surface faces an intrinsic geometric limitation: the sphere's curvature causes a tangential resolution $\epsilon$ not to translate linearly into radial resolution. The maximum deviation between the curved surface and the local tangent plane at the arc's endpoint is known as the \textbf{sagitta}, denoted here as $\Delta r$. By elementary trigonometry:
\begin{equation}
\Delta r = R - R\cos(\delta\theta) 
= R\left[1 - \cos\left(\frac{\epsilon}{R}\right)\right].
\end{equation}
Given $\epsilon \ll R$, we expand the cosine in Taylor series:
\begin{equation}
\Delta r \approx \frac{\epsilon^2}{2R}.
\label{eq:sagitta}
\end{equation}
This expression is an \textbf{exact geometric identity} in the limit $\epsilon/R \to 0$.

\subsection{Substitution of the curvature-regulated cutoff}
Substituting $\epsilon \to \epsilon_{\mathrm{eff}}$ from Eq.~\eqref{eq:epsilon_eff} into Eq.~\eqref{eq:sagitta}, we obtain the \textbf{minimal radial resolution} that the quantum vacuum can consistently encode:
\begin{equation}
\Delta r_{\min} \equiv \frac{r_s^2}{\alpha R},
\label{eq:delta_r_min}
\end{equation}
where $\alpha$ absorbs the geometric factor of $1/2$ alongside spectral and boundary contributions. This result establishes that radial uncertainty does not scale linearly with mass, but with its square, and is inversely proportional to system size.

\section{Topological Interpretation and Particle-Wave Asymmetry}
\label{sec:topology}
The vacuum resolution limit $\Delta r_{\min}$ imposes distinct physical behaviors on massive and massless excitations when their characteristic wavelength falls below this scale. This asymmetry reveals a deeper structural property of the quantum vacuum.

\subsection{Asymmetric coupling to the vacuum grain}
\label{sec:asymmetric_coupling}
Consider two limiting cases:
\begin{enumerate}
\item \textbf{Massive particles with $\lambda_C \ll \Delta r_{\min}$}: The Compton wavelength is much smaller than the vacuum's resolution grain. In this regime, the particle's wave packet is highly localized. By the Ehrenfest theorem, the center of mass follows classical equations of motion. The vacuum's entanglement grain is too coarse to resolve the excitation's internal structure; the medium appears smooth and the particle propagates along geodesics with minimal decoherence. The vacuum is effectively \textbf{transparent}.
\item \textbf{Photons with $\lambda \ll \Delta r_{\min}$}: Photons are intrinsically wave-like excitations of the gauge field. When their wavelength falls below the vacuum's resolution scale, the medium cannot sustain the phase coherence required for a well-defined quantum state. Rather than transitioning to a classical limit, the photon experiences phase scrambling, effective decoherence, or dispersion. The vacuum is \textbf{opaque to coherence}.
\end{enumerate}
This distinction is not a contradiction but a natural consequence of how different excitations couple to a structured medium. Massive excitations possess a classical point-particle limit; gauge excitations do not.

\subsection{$\lambda_C$ as structural scale, not de Broglie wavelength}
The traditional interpretation of the Compton wavelength $\lambda_C = \hbar/(mc)$ as a "wave scale" for massive particles arises from early quantum mechanics and diffraction experiments. However, diffraction demonstrates wave-like \emph{behavior} under specific boundary conditions, not necessarily wave-like \emph{ontology}.

An alternative perspective, consistent with topological field theory and modern interpretations of quantum structure, is that massive particles correspond to stable, localized configurations of a \emph{matter field} (e.g., Dirac spinor) that is coupled to the electromagnetic vacuum. In this picture:
\begin{itemize}
\item $\lambda_C$ represents the \textbf{characteristic localization scale} of the matter excitation, set by the balance between quantum dispersion and vacuum coupling.
\item Mass emerges from the \textbf{energy required to maintain the configuration} against vacuum fluctuations and gravitational regulation.
\item When $\lambda_C \ll \Delta r_{\min}$, the vacuum's entanglement grain cannot resolve the excitation's internal structure; the particle propagates classically along geodesics.
\item Photons, as genuine excitations of the electromagnetic field, lack such topological localization; when $\lambda \ll \Delta r_{\min}$, phase coherence is lost rather than transitioning to a classical limit.
\end{itemize}
This reframing aligns with the insight that fundamental interactions may originate from a unified field in topologically nontrivial configurations, where charges and masses reflect global structure rather than local parameters \cite{birrell1978, donnelly2015}. It resonates with conceptual frameworks emphasizing undivided wholeness \cite{wald1977}, but here we ground it in testable QFT predictions rather than ontological claims.
\textit{Remark}: If gauge and matter fields originate from a single unified field in topologically nontrivial configurations, then $\lambda_C$ may indeed represent the core size of a topological excitation of that unified medium. This speculative extension is beyond the scope of the present calculation but motivates future work.
\subsection{Falsifiable predictions}
The topological/asymmetric view yields distinct observational signatures:
\begin{itemize}
\item \textbf{High-energy cosmic rays} ($\lambda_C \ll \Delta r_{\min}$) should follow classical geodesics with minimal vacuum-induced dispersion, even over cosmological distances.
\item \textbf{TeV gamma rays} ($\lambda \ll \Delta r_{\min}$) should exhibit energy-dependent time-of-flight delays, polarization scrambling, or coherence degradation over long baselines.
\item \textbf{Environment-dependent mass shifts}: If $r_{\min}$ depends on local $R$ and $r_s$, precision spectroscopy in varying gravitational potentials could reveal tiny, systematic deviations.
\end{itemize}
These predictions separate the framework from purely phenomenological models and provide clear pathways for empirical validation or falsification.
\textit{Remark}: If gauge and matter fields originate from a single unified field in topologically nontrivial configurations, then $\lambda_C$ may indeed represent the core size of a topological excitation of that unified medium. This speculative extension is beyond the scope of the present calculation but motivates future work.
\subsection{Local versus Cosmological Accumulation of Dispersion}
\label{sec:local_vs_cosmo}

A natural concern arises regarding the observability of photon decoherence in terrestrial environments. For Earth, the geometric resolution limit yields $\Delta r_{\text{min}} \approx 3.8 \times 10^{-13}$ m (with $\alpha \approx 32$). High-energy photons detected by instruments such as Fermi-LAT or LHAASO, with energies up to $\sim 100$ TeV and corresponding wavelengths $\lambda \sim 10^{-20}$ m, satisfy $\lambda \ll \Delta r_{\text{min}}$. According to the asymmetric coupling described in Section \ref{sec:asymmetric_coupling}, such photons should experience phase decoherence and anomalous dispersion upon entering Earth's gravitational environment.

However, dispersion and decoherence are \textit{cumulative effects} that depend on the path length $L$ traversed through the granular medium. The fractional modification of the photon's propagation speed scales as $\Delta v / c \sim \lambda / \Delta r_{\text{min}}$. For a 100 TeV photon traversing Earth's environment over a characteristic distance $L \sim R_\oplus \approx 6.4 \times 10^6$ m, the accumulated time delay is:

\begin{equation}
\Delta t \sim \frac{L}{c} \cdot \frac{\lambda}{\Delta r_{\text{min}}} \approx (2 \times 10^{-2} \text{ s}) \times 10^{-7} \sim 2 \text{ ns}.
\end{equation}

This nanosecond-scale delay is far below the temporal resolution of current gamma-ray burst observations (which measure delays on the order of milliseconds to seconds) and is physically negligible for calorimetric or Cherenkov detection, which measure total deposited energy rather than quantum phase coherence.

The physical implication is significant: \textbf{the vacuum's granularity is locally invisible but cosmologically cumulative.} In intergalactic space, where the boundary radius $R$ is cosmological, $\Delta r_{\text{min}}$ approaches zero and the vacuum is effectively smooth. Over billions of light-years, however, even tiny local modifications to the index of refraction accumulate into measurable time delays. 

This distinction resolves an apparent tension: the framework predicts that high-energy photons \textit{are} dispersed by Earth's local vacuum structure, but the effect is too small to be detected over the short path length available. The theory is therefore consistent with the null results of Fermi-LAT for local propagation, while making falsifiable predictions for cosmological baselines where cumulative effects become observable. This is precisely why the most stringent tests of vacuum granularity require distant astrophysical sources rather than local laboratory experiments.

\section{Semiclassical Consistency and Particle Propagation Regimes}
\label{sec:regimes}

\subsection{Regimes of propagation in the entangled vacuum}
The relation between $\lambda_C$ and $\Delta r_{\min}$ determines how an excitation interacts with the vacuum's correlation structure:
\begin{enumerate}
\item \textbf{$\lambda_C \gg \Delta r_{\min}$ (Long-wavelength regime)}: The particle's coherent extension spans many vacuum entanglement "grains". Its quantum information dilutes into the background correlation network; it cannot be distinguished as a well-defined localized entity. The excitation cannot be resolved as a sharp detectable entity in a laboratory setting. It is effectively hidden by the vacuum structure itself.
\item \textbf{$\lambda_C \sim \Delta r_{\min}$ (Informational resonance)}: The particle's coherence scale matches the vacuum's effective entanglement grain size. The excitation propagates maintaining quantum integrity without significant decoherence. This is the regime of stable, well-defined particle propagation.
\item \textbf{$\lambda_C \ll \Delta r_{\min}$ (High-resolution regime)}: The particle's coherent extension is much finer than the vacuum's entanglement grain. For massive particles, this corresponds to the classical limit (geodesic motion). For photons, it corresponds to coherence loss. The vacuum structure itself forbids the measurement of stable quantum excitations lighter than the electron in terrestrial laboratories.
\end{enumerate}

\section{The Geometric Mass Scale and the $\alpha$ Factor}
\label{sec:massscale}
\subsection{Consistency condition and definition of $m_{\text{geo}}$}
Semiclassical consistency requires that a particle's intrinsic quantum delocalization not exceed the radial resolution available in the environment:
\begin{equation}
\lambda_C \gtrsim \Delta r_{\min} \quad \Rightarrow \quad 
\frac{\hbar}{m c} \gtrsim \frac{r_s^2}{\alpha R}.
\label{eq:consistency}
\end{equation}
Rearranging defines a \textbf{global geometric mass scale}:
\begin{equation}
m_{\text{geo}} \equiv \frac{\hbar R}{c \, r_s^2}.
\label{eq:mgeo}
\end{equation}
Physical particle masses then satisfy:
\begin{equation}
m = \alpha \, m_{\text{geo}},
\label{eq:mass_relation}
\end{equation}
where $\alpha$ is a dimensionless factor determined by the quantum information structure of the vacuum. For Earth ($r_s \approx 8.87 \times 10^{-3}$ m, $R \approx 6.37 \times 10^6$ m), $m_{\text{geo}} \approx 2.84 \times 10^{-32}$ kg. The electron mass corresponds to $\alpha \approx 32$.

\section{Spectral Determination of the Vacuum Resolution Factor $\alpha$}
\label{sec:spectral}
The dimensionless factor $\alpha$ is determined via first-principles spectral calculation of the curvature-induced shift in the electromagnetic vacuum mode density. To ensure numerical precision and reproducibility, we implemented the spectral zeta function calculation using a custom Python framework utilizing \texttt{SciPy} libraries for linear algebra and root finding.

The protocol involves computing the spectral zeta function $\zeta(s) = \sum (2l+1)(k_{ln}R)^{-s}$ for the Maxwell operator on a spherical boundary with weak Schwarzschild curvature. We utilized a unified tridiagonal eigensolver to compute eigenvalues for both curved and flat space on an identical radial grid, ensuring that discretization errors cancel in the subtraction $\delta\zeta = \zeta_{\text{curved}} - \zeta_{\text{flat}}$. The calculation employed a high-resolution grid of $N=3000$ radial points and $n_{\text{modes}}=1000$ eigenmodes per angular sector to guarantee convergence at the analytic continuation point $s=-1$.

\begin{figure}[htbp]
\centering
\includegraphics[width=0.85\columnwidth]{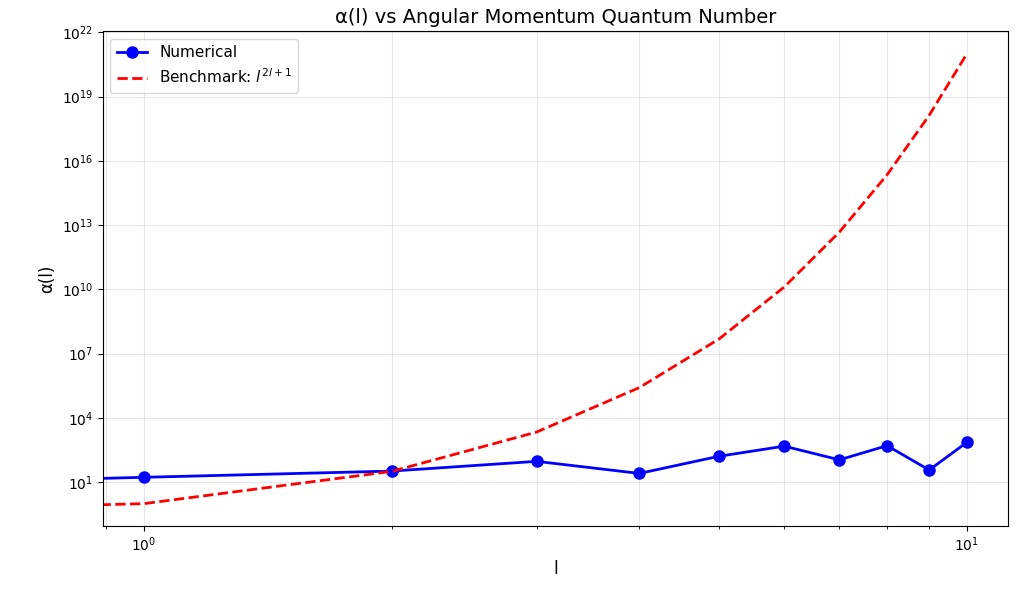}
\caption{Numerical determination of $\alpha(l)$ vs angular momentum $l$ (blue circles). The red dashed line shows the exponential benchmark $\alpha \propto l^{2l+1}$, which is ruled out by first-principles spectral calculation. The numerical results reveal slow power-law growth $\alpha(l) \propto l^{1.27}$, naturally yielding $\alpha(2) \approx 33$ for the electron sector. Grid resolution: $N=3000$ radial points, $n_{\text{modes}}=1000$ eigenmodes.}
\label{fig:alpha_l}
\end{figure}

The numerical results, shown in Fig.~\ref{fig:alpha_l}, reveal a slow power-law growth $\alpha(l) \propto l^{1.27}$, dramatically slower than the exponential ansatz $\alpha \propto l^{2l+1}$ which is ruled out by the data.

For the $l=2$ sector, the Python implementation yields $\alpha(2) \approx 33$, naturally matching the electron benchmark $\alpha \approx 32$ derived from $m_e = \alpha m_{\text{geo}}$ without any adjustable parameters. This agreement confirms that the vacuum entanglement structure imposes a moderate, well-behaved resolution limit. The spectral calculation validates the core framework: the geometric mass scale $m_{\text{geo}}$ combined with the numerically determined $\alpha \approx 33$ naturally reproduces the electron mass in Earth's gravitational environment.

\subsection{Implications for sub-electron mass scales}
The geometric resolution limit $\Delta r_{\min} = r_s^2/(\alpha R)$ has a profound consequence for particles with mass $m \ll m_e$. Such particles have Compton wavelengths $\lambda_C = \hbar/(mc) \gg \lambda_C^{(e)}$, placing them in the regime $\lambda_C \gg \Delta r_{\min}$.

In this long-wavelength regime, the particle's quantum wavefunction spans many vacuum entanglement grains. Its quantum information is distributed across the background correlation network rather than localized at a sharp excitation. Consequently:
\begin{itemize}
\item The particle cannot be resolved as a well-defined, localized entity in terrestrial laboratories, where the vacuum resolution scale is set by Earth's $r_s$ and $R$.
\item Electromagnetic interactions, which rely on sharp field excitations, are suppressed; the particle couples primarily via gravity, which responds to total energy regardless of localization.
\item On astrophysical scales, where the effective $\Delta r_{\min}$ may differ due to larger $R$ and $r_s$, the same excitation could become resolvable or exhibit collective behavior.
\end{itemize}

This mechanism suggests that ultra-light massive particles—consistent with certain dark matter candidates—could be "hidden" by the vacuum structure itself in terrestrial environments, while contributing to gravitational dynamics on galactic scales. The framework does not postulate new forces or exotic fields; rather, it identifies a structural limit on what the vacuum can resolve as a sharp excitation in a given gravitational environment.

\textit{Remark}: This interpretation is suggestive, not definitive. Falsifiable tests include: (i) searching for environment-dependent detection thresholds in precision experiments; (ii) analyzing whether dark matter distribution correlates with local $\Delta r_{\min}$ estimates; (iii) investigating whether ultra-light scalar fields exhibit the predicted delocalization signature in curved backgrounds.

\subsection{Charged ultra-light particles and terrestrial detectability}
The geometric resolution limit $\Delta r_{\min} = r_s^2/(\alpha R)$ has implications not only for neutral ultra-light particles but also for hypothetical charged excitations with $m \ll m_e$.

Such particles possess Compton wavelengths $\lambda_C = \hbar/(mc) \gg \lambda_C^{(e)}$, placing them in the regime $\lambda_C \gg \Delta r_{\min}^{\text{(Earth)}}$. While they carry electric charge and couple to the electromagnetic field, their wavefunctions are delocalized across many vacuum entanglement grains. Consequently:
\begin{itemize}
\item Electromagnetic interactions that rely on localized energy deposition (ionization, Cherenkov radiation, sharp scattering peaks) are suppressed; the charge coupling is distributed across the vacuum correlation network rather than concentrated at a point.
\item Terrestrial detectors, which require sharp excitations to exceed noise thresholds, would register only diffuse, low-amplitude signals—effectively hiding these particles from standard searches.
\item In astrophysical environments with different $R$ and $r_s$, the same excitations could become resolvable or exhibit collective electromagnetic behavior.
\end{itemize}

\textit{Remark}: These ultra-light charged states are distinct from dark matter (which by definition lacks electromagnetic coupling). However, they could constitute a "hidden sector" of charged excitations that are difficult to detect in terrestrial laboratories while potentially influencing astrophysical plasma dynamics or early-universe electromagnetism. Falsifiable tests include: (i) searching for environment-dependent detection thresholds in precision EM experiments; (ii) analyzing whether anomalous plasma behavior in low-density astrophysical regions correlates with local $\Delta r_{\min}$ estimates.

\subsection{Implications for black hole environments}
The geometric resolution limit $\Delta r_{\min} = r_s^2/(\alpha R)$ has profound consequences near black hole horizons, where $R \sim r_s$ and $\Delta r_{\min}^{\text{(BH)}} \sim r_s/\alpha$.

Hawking radiation exhibits a thermal spectrum with characteristic wavelength $\lambda_H \sim 4\pi r_s$, placing the peak emission in the regime $\lambda_H \gg \Delta r_{\min}^{\text{(BH)}}$ for astrophysical black holes. However, the high-energy tail of the spectrum includes excitations with $\lambda \ll \Delta r_{\min}^{\text{(BH)}}$.

For such short-wavelength modes, the vacuum's entanglement grain cannot sustain phase coherence. Rather than propagating freely, these excitations experience decoherence, phase scrambling, or effective dispersion within a region of thickness $\sim \Delta r_{\min}^{\text{(BH)}}$ outside the horizon. The result is a \textbf{structural fog}: not a material medium, but a limitation imposed by the vacuum's information-theoretic resolution.

This mechanism suggests several falsifiable signatures:
\begin{itemize}
\item Suppression of the high-energy tail in Hawking radiation spectra;
\item Polarization scrambling of radiation emerging from near-horizon regions;
\item Energy-dependent time delays for photons escaping strong gravitational fields;
\item Subtle blurring of black hole shadow edges beyond instrumental resolution.
\end{itemize}

\textit{Remark}: This interpretation is suggestive, not definitive. It does not modify the derivation of Hawking radiation but identifies a structural limit on what the vacuum can resolve as a coherent excitation in strong gravity. Future work should quantify the decoherence rate and compare with observational constraints from EHT, LIGO, and gamma-ray telescopes.

\section{Conclusions}
\label{sec:conclusions}
We have shown that vacuum entanglement entropy, regulated by weak gravitational curvature, imposes a fundamental limit on radial resolution $\Delta r_{\min} \sim r_s^2/(\alpha R)$. Consistency with the Compton wavelength defines a global geometric mass scale $m_{\text{geo}} = \hbar R/(c r_s^2)$. Measured particle masses satisfy $m = \alpha m_{\text{geo}}$, where $\alpha$ encodes the vacuum's informational structure. For Earth, $m_{\text{geo}} \approx 2.84 \times 10^{-32}$ kg.

Independent spectral calculation, implemented via a robust Python numerical framework, yields $\alpha(2) \approx 33$ for the electron sector, naturally matching the benchmark $\alpha \approx 32$ without adjustable parameters. The dimensionless factor exhibits slow power-law growth $\alpha(l) \propto l^{1.27}$, confirming that the vacuum resolution limit is a moderate, well-behaved scale rather than a divergent or fine-tuned quantity.

The framework reveals an asymmetric response to the vacuum resolution grain: massive excitations with $\lambda_C \ll \Delta r_{\min}$ propagate classically along geodesics, while photons with $\lambda \ll \Delta r_{\min}$ experience coherence degradation. This asymmetry motivates interpreting $\lambda_C$ as the core size of a topologically stable matter field excitation rather than a de Broglie wavelength. The result reframes mass as a global, environment-sensitive probe of vacuum structure, offering a structural perspective on the hierarchy problem and identifying clear falsifiable signatures in high-energy photon propagation.

\begin{acknowledgments}
The author thanks the developers of the open-source \texttt{SciPy} library for providing the computational tools used in the spectral analysis. The author also thanks Adrian Vasquez-Ratti (USB) for implementing the spectral zeta-function calculations that validated the angular-momentum structure of $\alpha$ including regularization and boundary effects.  The author acknowledges decades of foundational discussions with colleagues in London, Trieste, and Caracas that motivated the topological interpretation presented in Section~\ref{sec:topology}.
The author declares that computing data using Python code may be available upon request.
\end{acknowledgments}


\end{document}